# Diagonalization Matrix Method of Solving the First Basic Problem of Hidden Markov Model in Speech Recognition System

**R. Gnanajeyaraman**
**Department of IT, VMKV Engg College, Salem**
r.gnanajeyaraman@gmail.com

**G. Seenivasan**
**Department of IT, VMKV Engg College, Salem**
seenip2p@gmail.com

**ABSTRACT**. This paper proposes a computationally efficient method of solving evaluation problem of Hidden Markov Model (HMM) with a given set of discrete observation symbols, number of states and probability distribution matrices. The observation probability for a given HMM model is evaluated using an approach in which the probability evaluation is reduced to the problem of evaluating the product of matrices with different powers and formed out of state transition probabilities and observation probabilities. Finding powers of a matrix is done by using the computationally efficient diagonalization method thereby reducing the overall computational effort for evaluating the Evaluation problem of HMM. The proposed method is compared with the existing direct method. It is found that evaluating matrix power by diagnolisation method is more suitable than that of the direct, method.

## Introduction

Sequential pattern recognition is a special case of a statistical pattern recognition which is approached by the use of the Hidden Markov Model (HMM) [AK99]. In statistical pattern recognition supervised or unsupervised classification are made. Modern speech recognizer systems are based on HMMs. Although the HMMs originally emerged in the domain of speech recognition, other interesting fields are handwriting recognition, biological sequence analysis, face recognition etc [***97].

Among two broadly classified signal models such as deterministic and statistical signal models, the deterministic signal model characterizes the real word signal by estimating the amplitude, phase and spectral density of the signal [***06]. In the statistical signal modeling, the statistical properties of the real word signal are characterized. While carrying out the analysis of HMM three fundamental problems are encountered. They are:
1. The evaluation of the probability of a sequence of observation given a HMM model,
2. Determination of the best sequence of model states and
3. The adjustment of model parameters which maximizes the probability of a sequence of observations.

## 1. The Hidden Markov Model

In HMM, the states are hidden but each state at every instant of time generates an observation symbol with certain probability depending upon which observation symbol it has generated among a given set of observation symbols [Nil05]. In the weather model the climatic states at discrete instants of time are not known but the ice-cream sale index at each instant of time is known. This weather model can now be considered as the HMM based stochastic model. The urn-ball model is a classic example of Hidden Markov Model.

Consider N large box containing several urn's, each urn containing balls of M different colors (with colors given by red, blue, green, orange, yellow and violet). Let the entire set up be concealed behind a veil, and let a sequence of colored balls are popping out at discrete instants of time, $t = 1, 2, 3, \ldots\ldots\ldots T$ from the randomly chosen urns and are visible without any clue for the observer about which colored ball is popped out from which urn.

The M distinct colored balls and N urns are considered as observation symbols and states of HMM model respectively [PS06]. The sequence of colored balls emanating from the set up are observable and are considered as the observation vector $\bar{O}$. This observation vector $\bar{O}$ can be modeled as the observable output of an HMM system. Given appropriate values of N, M, (Transition probability matrix), B (Observation probability matrix), and $\pi$ (Initial probability matrix), the signal model HMM can be used as a generator to give an observation sequence $O = o_1, o_2, \ldots o_T$ (where each observation $o_t$ is one of the symbols from V, and T is number of observations in the sequence). For convenience we use the compact notation $\lambda = (A, B, \pi)$ to indicate the complete parameter set of the model [CS04].

## 2. The Evaluation Problem of HMM

The three basic problems of HMM are Evaluation, Decoding, Learning and the solutions for them are useful in developing an efficient HMM based statistical signal model for certain class of real word signals. They are also useful in applying developed HMM model for many real-word applications [DM96]. Even though there are three problems, this paper concentrates to find the solutions of the first basic problem which is described as follows:

- **Evaluation** problem: Given a HMM model $\lambda = (A, B, \pi)$ and a sequence of observations $\bar{O} = [o_1, o_2, o_3, \ldots o_T]$, what is the probability that the observations are generated by the model $\lambda$ i.e., $P(\bar{O}/\lambda)$ [KM92]

## 3. Evaluation of Observation Sequence Probability and Computational Effect

Evaluating the probability $P(\bar{O}/\lambda)$ of getting an observation sequence $\bar{O} = [o_1, o_2, o_3, \ldots o_T]$, for a given HMM model is to solve the first basic problem of HMM. Every observation sequence has a hidden state sequence $\bar{Q} = q_1, q_2, \ldots q_T$ responsible for the formation of it. Hence the probability of occurrence of observation sequence $\bar{O}$ can be computed from the joint probability of occurrence of both observation sequence and state sequence by summing such joint probabilities for all possible state sequences of $\bar{Q}$. [IS07]

The number of calculations involved to calculate $P(\bar{O}/\lambda)$, by direct definition is $2T.N^T$. Since at every t = 1, 2, 3 ….. T, there are N possible states [BS01]. Thus the number of possible state sequences would be $N^T$.

For every state sequence exactly (2T-1) number of multiplications are needed and thus to evaluate the total number of multiplications needed are $(2T-1)N^T$ and the total number of additions are $N^T - 1$. This computational effort is quite large even with smaller values of N and T.

## 4. Proposed Method

In this paper a new method is proposed to evaluate the probability of a given discrete observation sequence $\bar{O} = [o_1, o_2, o_3, \ldots o_T]$ and a given discrete HMM $\lambda = (A, B, \pi)$. $P(\bar{O}/\lambda)$ is obtained by calculating the sum of products for all possible state sequence. By considering all possible state sequences, the product element $a_{q_1,q_2} \cdot b_{q_2}(o_2)$ can be obtained by using the transition probability matrix $A_{N \times N} = \{a_{ij}\}$ and a specific column of observation probability matrix $B_{N \times M} = \{b_j(k)\}$.

Let $o_2$ be an observation symbol $v_k$ among the alphabet $\{v_k\}$ of $M$ symbols, then $b_{q_2}(o_2 = v_k)$ is the $k^{th}$ column of $B_{N \times M} = \{b_j(k)\}$. The product $a_{q_1,q_2} \cdot b_{q_2}(o_2)$ can be evaluated by multiplying each column of the matrix $A = \{a_{ij}\}$ with the corresponding element in the column vector $b_{q_2}(o_2 = v_k)$ resulting in the matrix $\{A_k\}$. Thus assuming each observation symbol $v_k$ occurring $l_k$ times, the equation can be written as

$$P(\bar{O}/\lambda) = \sum_Q \pi_{q_1} b_{q_1}(o_1) . A_1^{l_1} . A_2^{l_2} . A_3^{l_3} \ldots \ldots A_M^{l_M} \qquad (1)$$

with each matrix $A_k^{l_k}$ of $N \times N$ dimension and with $l_1 + l_2 + l_3 + \ldots \ldots l_M = T - 1$.

### *Evaluating matrix power by Diagonalization Matrix Method*

In order to reduce the number of multiplications and additions the proposed method of evaluating matrix power by Diagonalization is developed with the assumption that each matrix $A_k$ is a non-singular matrix

20

and is diagonalizable. The diagonalization can be used to compute the powers of the matrix. Let $P_k$ be an invertible matrix formed by taking the Eigen vectors of $A_k$ as its columns.

The diagonalization is given by the equation

$$D_k = P_k^{-1} A_k P_k \tag{2}$$

The diagonal matrix $D_k$ for the number of states N= 3 is of the form

$$D_k = \begin{bmatrix} \lambda_{k1} & 0 & 0 & 0 & 0 \\ 0 & \lambda_{k2} & 0 & 0 & 0 \\ 0 & 0 & \lambda_{k3} & 0 & 0 \\ 0 & 0 & 0 & \lambda_{k4} & 0 \\ 0 & 0 & 0 & 0 & \lambda_{k5} \end{bmatrix}$$

as the matrix product is associative,

$$A_k^{l_k} = P_k D_k^{l_k} P_k^{-1} \tag{3}$$

where the term $D_k^{l_k}$ is given by

$$D_k^{l_k} = \begin{bmatrix} \lambda_{k,1}^{l_k} & 0 & 0 & 0 & 0 \\ 0 & \lambda_{k,2}^{l_k} & 0 & 0 & 0 \\ 0 & 0 & \lambda_{k,3}^{l_k} & 0 & 0 \\ 0 & 0 & 0 & \lambda_{k4}^{l_k} & 0 \\ 0 & 0 & 0 & 0 & \lambda_{k5}^{l_k} \end{bmatrix}$$

Let $P_k$ and $P_k^{-1}$ are given by the square matrices such that the matrix power $A_k^{l_k}$ as per equation (3) is given by a square matrix. Given the HMM parameters $A$ and $B$, as given in the method 1, the matrices $\{A_k\}$ and their corresponding $P_k$, $P_k^{-1}$ and $D_k$ matrices can be computed a prior and stored for further computations.

The matrix power $A_k^{l_k}$ for each $k$ is calculated by using the above described procedure using diagonalization procedure and thus would require computational effort only for the evaluation of $D_k^{l_k}$ and the subsequent matrix multiplications in evaluating $P_k . D_k^{l_k} . P_k^{-1}$. Since $D_k^{l_k}$ is a diagonal matrix power, it would require at the maximum $N \times (l_k - 1)$ of multiplication for the evaluation of it. The total number of such computations for evaluating diagonal matrices for all the matrices $\{A_k\}$ would be $\sum_{k=1}^{M} N \times (l_k - 1)$.

The number of multiplications for evaluating each element in the matrix $P_k D_k^{l_k} P_k^{-1}$ is $N$ and hence $N \times N^2 = N^3$ multiplications are required to evaluate the matrix power $A_k$ apart from the calculation of diagonal matrix power. The total number of multiplications required to evaluate the product of matrix powers in equation (1) given by $A_1^{l_1} . A_2^{l_2} . A_3^{l_3} .......... A_M^{l_M}$ and is given by $\left( \sum_{k=1}^{M} N \times (l_k - 1) \right) + \left( N^3 (M-1) \right)$.

The first term in equation (1) is to be evaluated and is given by $\pi_{q_1} b_{q_1}(o_1)$. For a given $o_1 = v_k$ the term $b_{q_1}(o_1)$ is the $k^{th}$ column in the observation probability matrix $B$. To evaluate $\pi_{q_1} b_{q_1}(o_1)$ element wise multiplication is to be done between the selected column vector of the matrix and the initial probability matrix $\pi$ hence the number of multiplications to be done would be $N$. Each element of thus obtained product vector $\pi_{q_1} b_{q_1}(o_1)$ multiplies the corresponding row vector elements of the product matrix $A_1^{l_1} . A_2^{l_2} . A_3^{l_3} .......... A_M^{l_M}$. Thus an additional number of multiplications given by $N + N^2$ are to be performed as a part of solution for the evaluation of $P(\overline{O}/\lambda)$. Hence the total number of multiplications using the proposed approach would be $N + N^2 + \left( \sum_{k=1}^{M} N(l_k - 1) \right) + \left( N^3 (M-1) \right)$ which can be smaller especially when the observation vector length is larger.

After all the multiplications are computed, all the elements in the final matrix are added to obtain $P(\overline{O}/\lambda)$. Hence the number of additions required in the proposed method is quite less and are given by $\left(N^2(N-1)\right)M+\left(N^2-1\right)$ as shown in Table 1.

**Table 1. The number of multiplications and additions involved in various methods of evaluating the observation probability**

| Computations | Direct Method | Proposed Method (Diagonalization of matrix) |
|---|---|---|
| Number of Multiplications | $(2T-1)N^T$ | $N+N^2+\left(\sum_{k=1}^{M}N(l_k-1)\right)+\left(N^3(M-1)\right)$ |
| Number of Additions | $N^T-1$ | $\left(N^2(N-1)\right)M+\left(N^2-1\right)$ |

## 5. Experimental Results

Speech signals from Home Automation System are considered for calculating the Speech Recognition Accuracy by implementing the proposed two methods on the DSP 6713 Processor. In step1 evaluating matrix power by Diagonalization Matrix Method is implemented on the DSP Processor. Speech signals of constant time duration of 4 seconds are tested by various states. For each state the recognition accuracy is computed and recorded as shown in Table 2.

In step 2 Evaluating matrix power with Diagonalization method is implemented on the DSP Processor. Speech signals of constant time duration of 8 seconds are tested by various states. For each state the recognition accuracy is computed and recorded as shown in Table II. It shows the comparative study of results for the Direct and Diagnalization Matrix Method and from the readings the graphs are plotted as shown in figures 1 to 4. It shows that the Evaluating matrix power with Diagonalization method gives enhanced result than the Direct method. Maximum Speech recognition accuracy obtained is 96.8% on 16 states.

**Table 2. Recognition accuracy for various numbers of states**

| State | Recognition Accuracy (%) | | | |
|---|---|---|---|---|
| | 4 Seconds | | 8 Seconds | |
| | Direct Method | Diagonalization Method | Direct Method | Diagonalization Method |
| 10 | 72.4 | 94.8 | 74.6 | 95.6 |
| 16 | 73.0 | 95.6 | 75.8 | 96.8 |
| 32 | 72.4 | 94.0 | 73.4 | 95.0 |
| 64 | 71.0 | 93.0 | 71.6 | 93.8 |

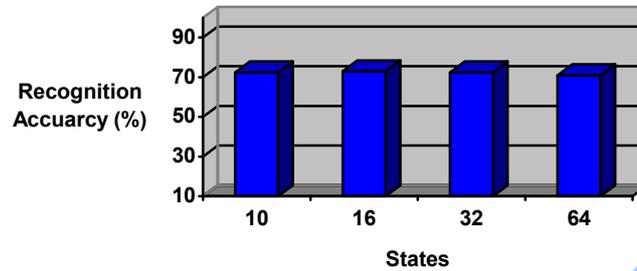

**Figure 1. Recognition accuracy for various states at 4 seconds duration in Direct Method**

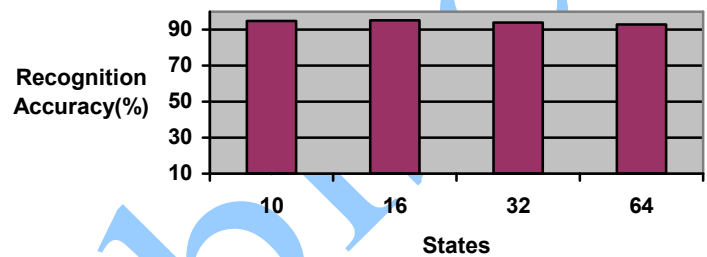

**Figure 2. Recognition accuracy for various states at 4 seconds duration in Diagonalization Matrix Method**

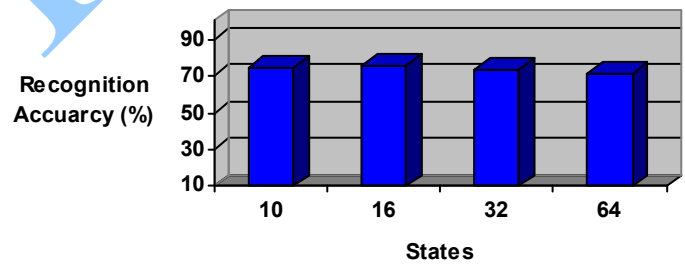

**Figure 3. Recognition accuracy for various states at 8 seconds duration in Direct Method**

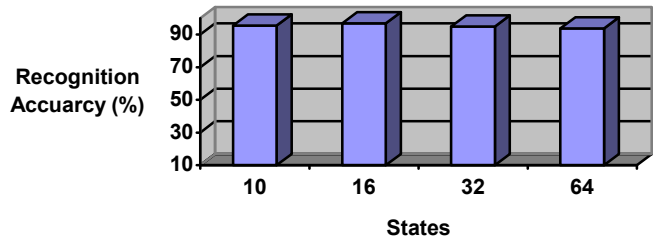

**Figure 4. Recognition accuracy for various states
at 8 seconds duration in Diagonalization Matrix Method**

## Conclusion

A computationally efficient method for solving the first of three basic problems of discrete HMM is presented in this paper. This paper presents a method for pre-computing a matrix for each distinct observation symbol in the available set of alphabet. With these pre-computed matrices available a forehand, evaluation problem of the HMM is reduced to the problem of evaluating the product of matrix powers. The evaluation of matrix powers can be done efficiently by using the procedure of the diagonalization of the matrix. The resulting product of matrix powers is element-wise multiplied with a vector obtained by considering the initial probability vector and the initial observation symbol. The overall computational effort is calculated and is compared with that of direct method of evaluating the observation probability. The Table I summarizes the number of multiplications and additions involved in the two methods of evaluating the observation probability. By implementing this new algorithm Speech Recognition accuracy is calculated on different states by using the Home automated speech signals. The accuracy obtained is 96.8% for 16 states test sample.

Any method for evaluating matrix power with a higher degree of computational efficiency can be used to substitute the method of diagonalization used in the proposed algorithm for evaluating the matrix power for further reduction in the computational effort.